\documentclass[twocolumn]{aastex62}

\definecolor{myblue}{RGB}{0, 100, 220}
\definecolor{myred}{RGB}{225, 0, 100}
\hypersetup{linkcolor=myblue,citecolor=myblue,filecolor=myblue,urlcolor=myblue}

\def\NV{\mbox{N\,{\sc v}}$\lambda 1240$}

\def\OIII{[\mbox{O\,{\sc iii}}]$\lambda 5007$}

\def\SII{[\mbox{S\,{\sc ii}}]$\lambda \lambda 6717,6731$}
\def\NII{[\mbox{N\,{\sc ii}}]$\lambda 6584$}
\def\NIIb{[\mbox{N\,{\sc ii}}]$\lambda 6584$}
\def\NIIa{[\mbox{N\,{\sc ii}}]$\lambda 6548$}

\def\Ha{{H$\alpha$}}
\def\Hb{{H$\beta$}}

\def\NIIHa{[\mbox{N\,{\sc ii}}]$\lambda 6583$/H$\alpha$}

\def\OIHa{[\mbox{O\,{\sc i}}]$\lambda 6300$/H$\alpha$}
\def\OIIIHb{[\mbox{O\,{\sc iii}}]$\lambda 5007$/H$\beta$}

\def\LOIIIs4{$L[\mbox{O\,{\sc iii}}]$/$\sigma^4$}

\def\ergs{${\rm erg}~{\rm s}^{-1}$}
\def\kms{${\rm km}~{\rm s}^{-1}$}
\newcommand{\ergcms}	{\ifmmode {\rm erg\,cm}^{-2}\,{\rm s}^{-1} \else erg\,cm$^{-2}$\,s$^{-1}$\fi}

\usepackage{gensymb}

\received{March 29, 2019}
\revised{May 29, 2019}
\accepted{June 8, 2019}

\shorttitle{\NIIHa\ vs. $\lambda_{\rm Edd}$ at $0.6<z<1.7$}
\shortauthors{Oh et al.}

\begin{document}

\title{An observational link between AGN Eddington ratio and \NIIHa\ at $0.6<z<1.7$}

\correspondingauthor{Kyuseok Oh}
\email{ohk@kusastro.kyoto-u.ac.jp}

\author[0000-0002-5037-951X]{Kyuseok Oh}
\altaffiliation{JSPS Fellow}
\affil{Department of Astronomy, Kyoto University, Kitashirakawa-Oiwake-cho, Sakyo-ku, Kyoto 606-8502, Japan}

\author{Yoshihiro Ueda}
\affiliation{Department of Astronomy, Kyoto University, Kitashirakawa-Oiwake-cho, Sakyo-ku, Kyoto 606-8502, Japan}

\author{Masayuki Akiyama}
\affiliation{Astronomical Institute, Tohoku University, Aramaki, Aoba-ku, Sendai, Miyagi 980-8578, Japan}

\author{Hyewon Suh}
\altaffiliation{Subaru Fellow}
\affiliation{Subaru Telescope, National Astronomical Observatory of Japan (NAOJ), National Institutes of Natural Sciences (NINS), 650 North A'ohoku place, Hilo, HI 96720, USA}

\author{Michael J. Koss}
\affiliation{Eureka Scientific, 2452 Delmer Street Suite 100, Oakland, CA 94602-3017, USA}

\author{Daichi Kashino}
\affiliation{Department of Physics, ETH Z\"{u}rich, Wolfgang-Pauli-Strasse 27, CH-8093, Z\"{u}rich, Switzerland}

\author{G\"{u}nther Hasinger}
\affiliation{European Space Astronomy Centre (ESA/ESAC), Director of Science, E-28691 Villanueva de la Ca\~{n}ada, Madrid, Spain}








\begin{abstract}

We present an observed relationship between Eddington ratio ($\lambda_{\rm Edd}$) and optical narrow-emission-line ratio (\NIIHa) of X-ray-selected broad-line active galactic nuclei (AGN) at $0.6<z<1.7$. We use 27 near-infrared spectra from the Fiber Multi-Object Spectrograph along with 26 sources from the literature. We show that the $\lambda_{\rm Edd}$ and \NIIHa\ ratio at $0.6<z<1.7$ exhibits a similar anti-correlation distribution of $\lambda_{\rm Edd}-$\NIIHa\ as has been found for local ($ \langle z \rangle = 0.036$), ultra-hard X-ray selected AGN. The observed distribution implies that there is a consistent relationship from local to $z\sim1.7$ which corresponds from the present time to 4 Gyr old. Further study of high redshift low Eddington ratio AGN (${\rm log}\lambda_{\rm Edd}<-2$) is necessary to determine fully whether the $\lambda_{\rm Edd}-$\NIIHa\ anti-correlation still holds in high-redshift AGN at low Eddington ratios.

\end{abstract}



\section{Introduction} \label{sec:intro}

Nebular emission lines play an important role in modern astronomy. Forbidden emission lines (e.g., \OIII, \NII, and \SII) and Balmer lines (\Ha, \Hb) in the optical band are frequently used to diagnose the physical state of active galactic nuclei (AGN) and star-forming activity \citep{Baldwin81, Veilleux87, kewley01, Schawinski07} and applied to distinguish central nuclear activity from star-formation. In AGN, higher ratios of collisionally excited forbidden lines are predicted because AGN produce more high energy photons than star-forming regions in a galaxy, which produce photoionization-induced Balmer emission lines. 
Both forbidden lines and Balmer lines are easy to detect as they are prominent in AGN hosts and star-forming galaxies. Due to being closely located in wavelength these lines suffer similar extinction and their ratios can therefore be used independent of extinction. In the framework of AGN unified model \citep{Antonucci93, Urry95}, furthermore, these prominent narrow emission-lines are extended on kpc scales and thus are visible regardless of accretion rate and viewing angle. Despite its efficiency as a useful diagnostic tool, several studies have reported that heavily obscured or dusty AGN and those with significant star formation may not be diagnosed as AGN using emission-line spectroscopy since obscuration and star formation can dilute emission-line strengths originated from AGN \citep{Elvis81, Iwasawa93, Griffiths95, Barger01, Comastri02, Rigby06, Caccianiga07, Trump15, Koss17}. 

Massive spectroscopic surveys such as Sloan Digital Sky Survey \citep{York00} have made it possible to explore emission-line properties, diagnostics, and Eddington ratios ($\lambda_{\rm Edd}$) of sizeable sample of local AGN. For example, \citet{Kewley06} showed that 85224 local emission-line galaxies selected from the SDSS at $0.04 < z < 0.10$ have a positive correlation between $\lambda_{\rm Edd}$ and a distance from the LINER regime in the \OIIIHb\ versus \OIHa\ diagram (see Fig.19 in \citealt{Kewley06}). Later, \citet{Stern13} presented emission-line diagnostics of 3175 nearby broad-line AGN from the SDSS ($z<0.31$) and found a dependence on $\lambda_{\rm Edd}$. 

Over the last decade new facilities in high-energy astrophysics such as \textit{INTEGRAL} \citep{Winkler03}, \textit{Swift} \citep{Gehrels04}, and \textit{NuSTAR} \citep{Harrison13} and their observations of AGN have greatly helped to explore the nature of obscured AGN even including Compton-thick sources (e.g., \citealt{Ricci15, Marchesi18}). High energy photons ($>10$ keV) are only biased against Compton-thick levels of obscuration at ${\rm log}N_{\rm H}>24.5$ ${\rm cm}^{-2}$ and are not subject to host galaxy contamination or dust obscuration \citep{Koss16}. Since 2004 December, in particular, the Burst Alert Telescope (BAT, \citealt{Barthelmy05}) on the \textit{Swift} satellite \citep{Gehrels04} has been successfully carrying all-sky hard X-ray observation ($14-195$ keV), providing monthly lightcurves as well as X-ray spectra \citep{Markwardt05, Tueller08, Tueller10, Baumgartner13, Oh18}. The most recent publication of \textit{Swift}-BAT all-sky hard X-ray survey\footnote{http://heasarc.gsfc.nasa.gov/docs/swift/results/bs105mon/} reached a sensitivity of $8.40\times10^{-12}$ $\ergcms$ over 90\% of the sky, identifying 1632 X-ray sources, of which 1099 are AGN including beamed sources \citep{Oh18}.

Over the past 5 years a massive effort has been made in investigating complete census of black hole masses ($M_{\rm BH}$) and accretion rates normalized by $M_{\rm BH}$ ($\lambda_{\rm Edd}$, $\lambda_{\rm Edd} \equiv L_{\rm bol}/L_{\rm Edd}$, where $L_{\rm Edd} \equiv 1.3 \times 10^{38} (M_{\rm BH}/M_{\odot})$) of X-ray selected AGN. Based on the 70-month catalog of \textit{Swift}-BAT \citep{Baumgartner13}, which identified 836 AGN, the BAT AGN Spectroscopic Survey Data Release 1 (BASS-DR1\footnote{http://bass-survey.com}, \citealt{Koss17, Lamperti17, Ricci17}) published $M_{\rm BH}$, $\lambda_{\rm Edd}$, narrow/broad emission-line strength, emission-line diagnostics, stellar velocity dispersion, and X-ray spectral properties for 642 AGN ($ \langle z \rangle \sim 0.05$) using various public surveys (the Sloan Digital Sky Survey and the 6dF Galaxy Survey; \citealt{Abazajian09, Jones09, Alam15}) and dedicated follow-up optical/near-infrared spectroscopic observations. 

\begin{figure}
	\includegraphics[width=\linewidth]{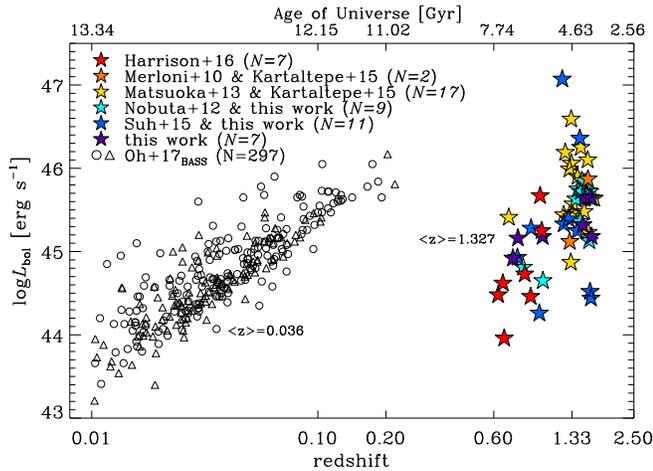} 
    \caption{Bolometric luminosity as a function of redshift. Empty circles (broad-line AGN) and triangles (narrow-line AGN) in black are local hard X-ray selected AGN from the BASS-DR1 \citep{Koss17, Oh17}. The high-redshift ($\langle z \rangle \sim 1.3$) X-ray-selected broad-line AGN used in this work are shown with color-filled stars. 
    }
    \label{fig:z_dist}
\end{figure}

Using optical emission-line strengths, bolometric luminosity ($L_{\rm bol}$), and $M_{\rm BH}$ for $\sim 300$ local AGN ($\langle z \rangle \sim 0.036$, Fig~\ref{fig:z_dist}), \citet{Oh17} showed that the \NIIHa\ ratio exhibits a significant anti-correlation with $\lambda_{\rm Edd}$ ($R_{\rm Pear}=-0.44$, $p$-${\rm value}= 3\times10^{-13}$, root-mean-square deviation $\sigma=0.28$ dex). The authors confirmed that the reported anti-correlation still holds good in both broad-line and narrow-line AGN, and it is free from different apertures affects that applied for the heterogeneous observations in the BASS-DR1. With a measured anti-correlation with 0.28 dex of $\sigma$ the \NIIHa\ ratio could, in principle, be used to infer accretion efficiencies of high-redshift obscured AGN.  

At high redshifts, X-ray observations provide the most complete and numerous samples of AGN including obscured, unobscured, and lower luminosity AGN. For instance, the deepest \textit{Chandra} surveys can reach AGN that are 100 times less luminous than wide field optical surveys where AGN are 500 times more numerous \citep{Brandt10}. While the X-rays provide a reliable tracer of bolometric luminosity up to Compton-thick levels, the black hole mass and Eddington ratio require further observations. For distant obscured AGN, the bulge velocity dispersion, which can be turned into $M_{\rm BH}$ and eventually $\lambda_{\rm Edd}$ assuming $M_{\rm BH}$ - $\sigma_{*}$ relation holds at high redshift, is extremely challenging to measure at high S/N to estimate a black hole mass due to the faintness of the host galaxy.

In this paper, we investigate the observed relationship between $\lambda_{\rm Edd}$ and the narrow-emission-line ratio, \NIIHa, using 53 X-ray selected broad-line AGN at $0.6 < z < 1.7$, extending the previous study of \citet{Oh17} beyond the local Universe. This paper is organized as follows. In Section 2, we describe the samples we collated from the literature, and present the spectral line fitting results. In Section 3, we show the observed relationship between $\lambda_{\rm Edd}$ and \NIIHa\ ratio at $0.6 < z < 1.7$. Finally, we discuss our findings and summarize our result in Section 4, with caveats and cautions. We assume a cosmology with $h=0.70$, $\Omega_{\rm M}=0.30$, and $\Omega_{\Lambda}=0.70$ throughout this work.

\begin{deluxetable*}{lcccc}[ht]
\tablecaption{Summary of Samples \label{tab:samples}}
\tablewidth{0pt}
\tablehead{
\colhead{Literature / Survey\tablenotemark{a}} & 
\colhead{N\tablenotemark{b}} & 
\colhead{${\rm Inst.}_{L_{\rm bol}}$\tablenotemark{c}} &
\colhead{${\rm Inst.}_{M_{\rm BH}}$\tablenotemark{d}} & 
\colhead{${\rm Inst.}_{[\mbox{N\,{\sc ii}}]~\lambda 6583/H\alpha}$\tablenotemark{e}} 
}
\startdata
\citealt{Harrison16}	 / KASH$z$					& 	7	& 	\textit{XMM-Newton}, CDF-S	& \textit{VLT}/KMOS 		 	 &	\textit{VLT}/KMOS										\\
	\citealt{Merloni10, Kartaltepe15} / FMOS-COSMOS	& 	2	& 	CCLS						& \textit{VLT}/VIMOS		 	 &	\textit{Subaru}/FMOS									\\
	\citealt{Matsuoka13,Kartaltepe15} / FMOS-COSMOS	& 	17	& 	CCLS						& \textit{Subaru}/FMOS		 &	\textit{Subaru}/FMOS									\\
	\citealt{Nobuta12} / SXDS						&	9	& 	\textit{XMM-Newton}			& \textit{Subaru}/FMOS		 &	\textit{Subaru}/FMOS									\\
	\citealt{Suh15}	/ FMOS-COSMOS					& 	11	& 	CDF-S, \textit{XMM}-LH		& \textit{Subaru}/FMOS		 &	\textit{Subaru}/FMOS									\\
	Suh et al. \textit{(in prep.)} / FMOS-COSMOS		&	7	& 	CCLS						& \textit{Subaru}/FMOS		 &	\textit{Subaru}/FMOS									\\
\enddata
\tablenotetext{a}{Source of literature with name of the survey.}
\tablenotetext{b}{Size of sample.}
\tablenotetext{c}{Telescope or survey used to measure 2-10 keV luminosity.}
\tablenotetext{d}{Telescope/instrument used to measure $M_{\rm BH}$.}
\tablenotetext{e}{Telescope/instrument used to measure \NIIHa\ narrow-emission-line ratio.}
\end{deluxetable*}

\section{Sample Selection, Data, and Measurements}
\label{sec:data}
In order to investigate the observed link between $\lambda_{\rm Edd}$ and \NIIHa\ at higher redshifts ($\langle z \rangle \sim 1.3$) to directly compare it with the previous study using ultra-hard X-ray selected AGN in the local Universe ($\langle z \rangle \sim 0.036$, \citealt{Oh17}), we constructed our samples from literature as follows: 
\begin{enumerate}
 \item Unobscured AGN that have spectroscopically confirmed redshifts featuring broad-emission-lines.
 \item X-ray selected AGN whose intrinsic rest-frame 2-10 keV luminosities are available.
 \item Those with deblended narrow-emission-line profiles in \NII\ and \Ha.
\end{enumerate}

We briefly describe the literature used in our investigation in the following section.

\subsection{The Subaru / \textit{XMM-Newton} Deep Survey}
\label{sec:sxds}
The Subaru / \textit{XMM-Newton} Deep Survey (SXDS) \citep{Sekiguchi05} is one of the largest, multiwavelength survey that combines dedicated spectroscopic observations from the far-UV to mid-IR \citep{Nobuta12, Akiyama15} with soft ($0.5-2$ keV) and hard ($2-10$ keV) band X-ray observations \citep{Ueda08}. The SXDS field (R.A.: $02^{\rm h}18^{\rm m}$, decl.: $-05$\degree) was observed with \textit{XMM-Newton} using 100 ks of exposure for a central 30\arcmin\ diameter field and 50 ks for 6 flanking fields.  Out of 896 broad-line AGN candidates observed in the SXDS survey, 851 sources were observed using the Fiber Multi-Object Spectrograph (FMOS) on the Subaru telescope \citep{Kimura10}. The FMOS was used to observe the wavelength between 9000\AA\ and 18000\AA\ with a spectral resolution of $R\sim800$ at $\lambda\sim1.55\micron$ in the low-resolution mode. Due to the faintness of the targets, 586 optical and near-infrared spectra of the total 896 sources were observed \citep{Nobuta12}. 

In this work, we performed spectral line fitting for 84 sources out of these 586, whose broad \Ha\ components as well as \NII\ and narrow \Ha\ lines are measurable at $0.63<z<1.70$. We used {\tt gandalf} IDL code \citep{Sarzi06}, which was originally developed for spectral line decomposition of the 72 early-type galaxies of full SAURON sample \citep{Bacon01} and was applied later to more than 660,000 nearby SDSS narrow emission-line galaxies \citep{Oh11}\footnote{http://gem.yonsei.ac.kr/ossy/}. Following the application of {\tt gandalf} procedures in \citet{Oh15}, we first deredshifted FMOS spectra taking into account Galactic foreground extinction \citep{Schlafly11}. We then fitted the wavelength region of interest ($6000$\AA$-6750$\AA) focusing on deblending \NIIb\ and narrow \Ha\ emission-lines along with broad \Ha\ component using Gaussian emission-line templates. The \NIIa\ and \NIIb\ lines were tied each other with a pair of Gaussians holding the theoretical ratio of 2.96. Gaussian line widths were set as free parameters with upper limits of 600 \kms\ for the narrow-lines and 10,000 \kms\ for the broad lines, respectively. When necessary, multiple Gaussian components were applied to fit broad \Ha\ components following established procedures \citep{Collin06, LaMura07, Mullaney08}. After fitting, 9 reliable fits showing deblended \NII\ and \Ha\ narrow emission-lines along with a broad \Ha\ component (Fig.~\ref{fig:fit_N12O18}) were found. The majority of the remaining sources show low signal-to-noise ratio or dominant broad \Ha\ emission features with indistinguishable narrow emission-lines. 

Black hole mass measurements relying on single-epoch spectra are commonly used for broad-line AGN \citep{Kaspi00, McLure02, Vestergaard02, Woo02, Greene05, Shen08, Vestergaard09, Shen12, Trakhtenbrot12, Matsuoka13}. Virial black hole mass estimation assumes that gravity of a central black hole governs the kinematics of broad-line region (BLR), which yields the relationship between black hole mass, gas velocity, and radius of the BLR ($M_{\rm BH} \propto R\Delta V^{2}/G$). Broad Balmer lines (typically \Hb\ and \Ha) are used as a proxy of the average gas velocity, and the $R_{\rm BLR}-L$ relation obtained through reverberation mapping technique \citep{Kaspi00, Kaspi05, Bentz06, Bentz13} provides the radius of the BLR. We estimated the black hole mass of broad-line AGN using the broad \Ha\ luminosity and line width following \citet{Greene05}. 

We estimated the bolometric luminosities of broad-line AGN using the intrinsic $2-10$ keV rest-frame luminosity provided by \citet{Nobuta12} and the median bolometric correction ($k=20$) as measured by \citet{Vasudevan09}. We applied the same bolometric correction to the same energy band ($2-10$ keV) following that of \citet{Oh17}. We calculated $\lambda_{\rm Edd}$ ($L_{\rm Edd} \equiv L_{\rm bol}/L_{\rm Edd}$) by combining the estimated $L_{\rm bol}$ and $L_{\rm Edd}$, assuming $L_{\rm Edd} \equiv 1.3 \times 10^{38} (M_{\rm BH}/M_{\odot})$. A summary of the samples is provided in Tab.~\ref{tab:samples} with measured quantities provided in Tab.~\ref{tab:data}.

\subsection{The FMOS / X-ray surveys}
The \textit{Chandra COSMOS Legacy} Survey (CCLS, \citealt{Civano16}) is a 2.2 ${\rm deg}^{2}$ \textit{Chandra} survey of the COSMOS field with a total exposure time of 4.6 Ms. The CCLS point source catalog provides X-ray measurements for 4016 X-ray point sources and their $2-10$ keV fluxes. Spectroscopic follow-up observations of the CCLS field have been carried in optical with DEIMOS \citep{Hasinger18} and in NIR with FMOS (\citealt{Silverman15, Kashino19}, Suh et al. in prep.). In this field, we employ three samples using NIR spectroscopy with FMOS. 

The first used the low-resolution mode of FMOS, which covers from 9000\AA\ to 18000\AA\ with a dispersion of $\sim5\AA$ ${\rm pixel}^{-1}$. \citet{Kartaltepe15} obtained 119 near-infrared spectroscopic data including \NIIHa\ emission-line ratio selected from the COSMOS field at $z<1.7$. \citet{Matsuoka13} also obtained 43 near-infrared spectra of moderate-luminosity broad-line AGN at $z<1.8$ found in the COSMOS and Extended Chandra Deep Field-South Survey, using the low-resolution mode of FMOS instrument. Another effort on measuring $M_{\rm BH}$ of 89 broad-line AGN in the COSMOS field at $1<z<2.2$ had been made by \citet{Merloni10} using the VIMOS multi-object spectrograph on the European Southern Observatory's Very Large Telescope (ESO-VLT) and virial black hole masses derived from MgII line width and continuum luminosity ($L_{\rm 3000}$) \citep{McGill08} were made. Although these archival spectra provide at least several dozen high redshift unobscured AGN samples with key quantities such as $M_{\rm BH}$ measured from the broad-line (either \Ha\ or MgII) and bolometric luminosity from the $2-10$ keV flux, however, the total number of usable sources in such a redshift range ($z<\sim 2$) is limited due to the availability of the \NII\ and \Ha\ line ratio ($N=19$). 

In addition to these 19 sources, we obtained 18 sources from the X-ray selected broad-line AGN sample of FMOS-COSMOS survey (\citealt{Schulze18}, Suh et al. in prep.), at redshifts $0.73 < z < 1.63$. The FMOS high-resolution mode observations cover four spectral regions (J-short: $0.92 - 1.12\micron$, J-long: $1.11 - 1.35\micron$, H-short: $1.40 - 1.60\micron$, H-long: $1.60 - 1.80\micron$) with a spectral resolution of R $\sim 2600$. More details of the survey design, instrumental performance, and observations, can be found in \citet{Silverman15}.

The bolometric luminosity and Eddington ratio of these sources were estimated using available X-ray catalogues such as CCLS \citep{Civano16}, \textit{Chandra} Deep Field South (CDF-S) \citep{Xue11}, Extended \textit{Chandra} Deep Field South (E-CDF-S) \citep{Lehmer05}, and \textit{XMM-Newton}-Lockman Hole (\textit{XMM}-LH) \citep{Brunner08}. In order to estimate bolometric luminosity ($L_{\rm bol}$), we applied a bolometric correction ($k=20$) to an absorption-corrected rest-frame $2-10$ keV luminosity. 

We disentangled the broad \Ha\ and narrow-lines for 18 sources from the FMOS-COSMOS survey (\citealt{Silverman15}, Suh et al. in prep.) by applying the spectral line fitting as same as that of the SXDS data shown in the Section~\ref{sec:sxds}. The achieved spectral line fit for the 18 sources are shown in Fig.~\ref{fig:fit_S18O18} and Fig.~\ref{fig:fit_S15O18}. By adopting single-epoch virial mass estimation recipe of \citet[equation 9]{Greene05}, we computed the $M_{\rm BH}$ from the broad \Ha\ line widths and luminosities. 

\subsection{KASH$z$ AGN survey}
The KMOS ($K$-band Multi-Object Spectrograph) AGN Survey at High redshift (KASH$z$) provides measurements of the broad \Ha, narrow \Ha, and \NII\ emission lines as well as $2 - 10$ keV measurements for 7 broad-line AGN at $0.63<z<0.98$ \citep{Harrison16}. The authors compiled the X-ray measurements for those 7 broad-line AGN from the 4 Ms CDFS catalogue \citep{Xue11} and 400 ks SXDS catalog \citep{Ueda08}. 

\section{\NIIHa\ and Eddington ratio relation at $0.6<z<1.7$}
\label{sec:ratio}

\begin{figure*}
\centering
	\includegraphics[width=0.68\textwidth]{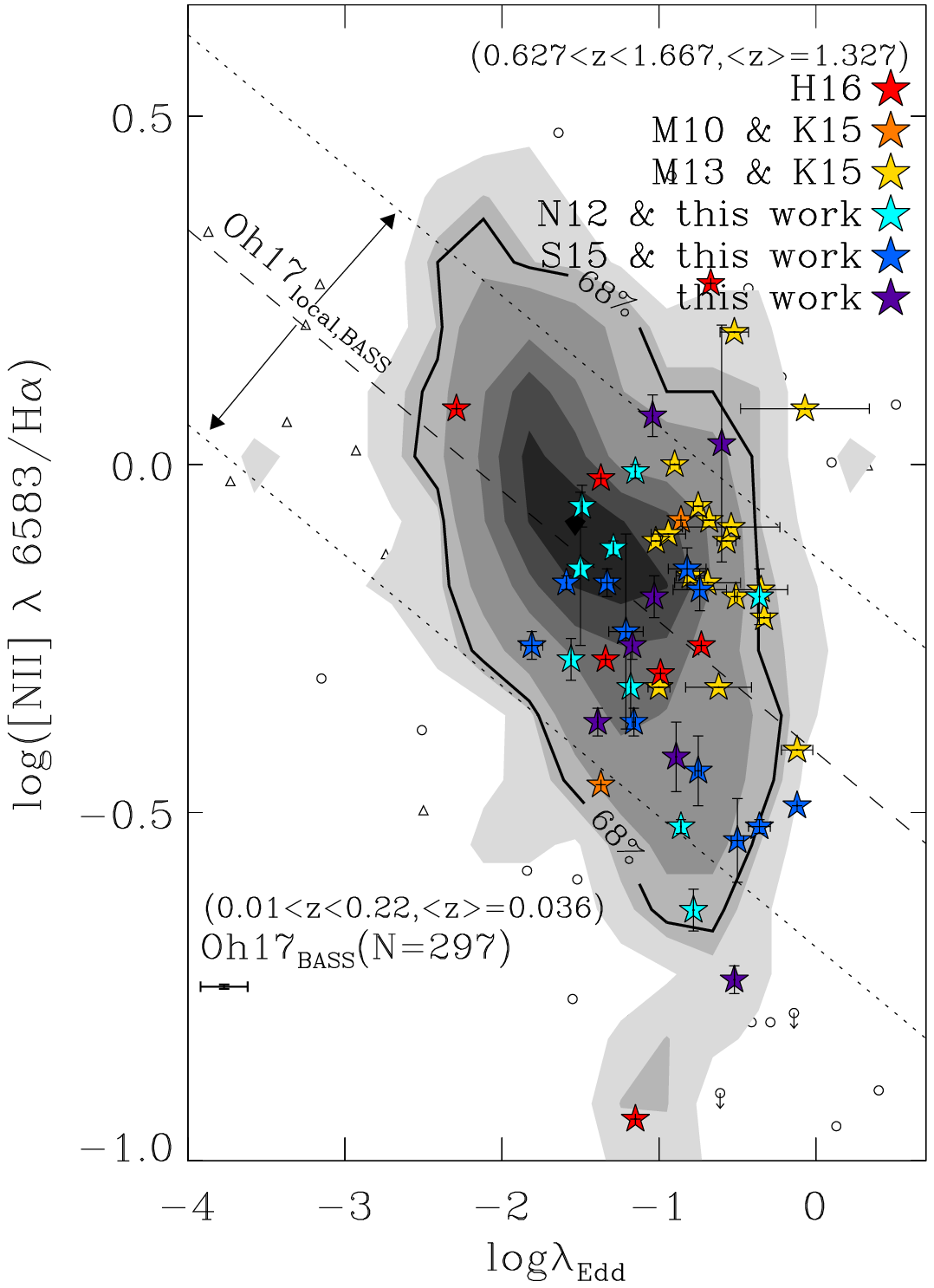}
    \caption{\NIIHa\ versus $\lambda_{\rm Edd}$ diagram for local and high-z X-ray selected broad-line AGN. The \NIIHa\ vs. $\lambda_{\rm Edd}$ relationship for local ($\langle z \rangle \sim 0.036$) hard X-ray selected AGN from \citet{Oh17} is shown with dashed-line, along with the rms deviation ($0.28$ dex, dotted-lines) and filled contours. The Pearson correlation coefficient and $p$-value are $-0.44$ and $3\times10^{-13}$, respectively. X-ray selected broad-line AGN at $0.626<z<1.667$ are presented with color-filled stars. When available, the measurements errors in both abscissa and ordinate are presented.}
    \label{fig:ratio}
\end{figure*}

We show the distribution of narrow \NIIHa\ as a function of $\lambda_{\rm Edd}$ for the 53 X-ray selected broad-line AGN at $0.6<z<1.7$ in Fig.~\ref{fig:ratio}. Local hard X-ray selected AGN at $\langle z \rangle \sim 0.036$ \citep{Oh17} are also shown with the filled contours along with the measured Bayesian linear regression fit (thin dashed line, equation~\ref{eq:regression}). Using 297 local AGN, \citet{Oh17} reported the statistically significant anti-correlation with $(-1.48\pm0.04, -0.98\pm0.13)$ of ($\alpha, \beta$), $-0.44$ of the Pearson $R$ coefficient, $3\times10^{-13}$ of $p$-value, and 0.28 dex of the root-mean-square (rms) deviation. 
\begin{equation}
	  \log({\rm [\mbox{N\,{\sc ii}}]~\lambda 6583}/{\rm {H\alpha}})= \alpha + \beta\log\lambda_{\rm Edd} 	
\label{eq:regression}
\end{equation} 

We find that the distribution of \NIIHa\ as a function of $\lambda_{\rm Edd}$ at $0.6<z<1.7$ is approximately similar with that of local AGN, mainly occupying high $\lambda_{\rm Edd}$ regime with $-0.86$ of median ${\rm log}\lambda_{\rm Edd}$. Unlike local AGN, statistical significance of the linear regression analysis for these high-z AGN is too poor to draw any definite relation at this time ($R_{\rm Pear}=-0.14$, $p$-${\rm value}=0.54$) because of a lack of low Eddington ratio sources. 

Confirming that the \NIIHa\ vs. $\lambda_{\rm Edd}$ relation still holds true for high-z AGN, as is hinted in Fig.~\ref{fig:ratio}, requires a much larger number of low $\lambda_{\rm Edd}$ AGN at ${\rm log}\lambda_{\rm Edd}<-2$. Indeed, \citet{Suh15} reported that broad-line AGN at $1<z<2.2$ showing ${\rm log}\lambda_{\rm Edd}<-2$ are very rare from their Subaru/FMOS observation for X-ray selected AGN along with previously published literature \citep{Gavignaud08, Merloni10, Shen11, Nobuta12, Matsuoka13}, as seen in Fig~\ref{fig:ratio}. Therefore, future high-resolution near-infrared spectroscopic observations for a statistically sizeable sample of low $\lambda_{\rm Edd}$ AGN at ${\rm log}\lambda_{\rm Edd} < -2$ are required to confirm the correlation.

We perform the Kolmogorov-Smirnov test (K$-$S test) for \NIIHa\ using sub-sample of local and high-z AGN based on the same range of ${\rm log}\lambda_{\rm Edd}$ ($>-1.81$). We find that \NIIHa\ distribution of local and high-z AGN are consistent with being drawn from the same underlying probability distribution with 0.11 of D statistic and 0.66 of $p$-value. We therefore find that while we cannot find a statistically significant linear correlation within the current dataset, the overall distribution of values is consistent with those at low redshifts.

\begin{figure*}[ht]
	\includegraphics[width=1.0\textwidth]{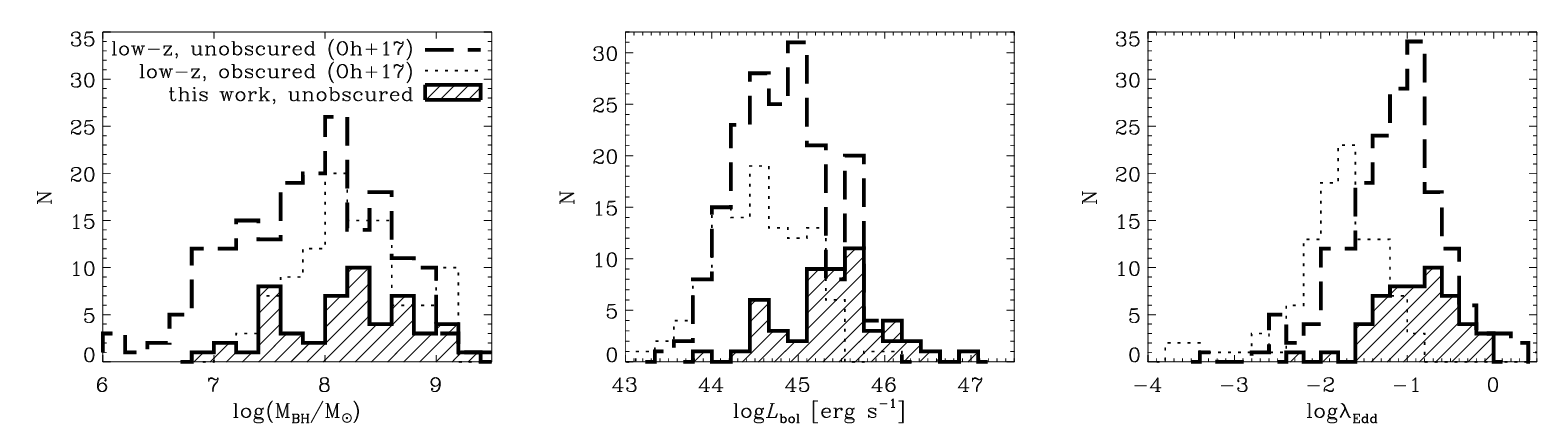}
    \caption{Histograms of ${\rm log}(M_{\rm BH}/M_{\rm \odot})$ (left-hand panel), ${\rm log}L_{\rm bol}$ (middle), and ${\rm log}\lambda_{\rm Edd}$ (right) for high-z AGN (hatched area), local obscured (narrow-line) AGN (dotted lines), and local unobscured (broad-line) AGN (long-dashed lines).}
    \label{fig:histograms}
\end{figure*}

Our sample of unobscured (broad-line) AGN covers the black hole mass range $6.9<{\rm log}(M_{\rm BH}/M_{\odot})<9.3$ and the bolometric luminosity range $44.0<{\rm log}L_{\rm bol}<47.1$ with the Eddington ratio ranging from $-2.29$ to $-0.07$ (Fig.~\ref{fig:histograms}). Compared to the local AGN shown in Fig.~\ref{fig:histograms}, the $M_{\rm BH}$ distribution of high-z AGN is not particularly different (0.11 of $p$-value from the K$-$S test) while the $L_{\rm bol}$ ($3.1\times10^{-11}$ of $p$-value) and their resultant derivative, $\lambda_{\rm Edd}$ ($3.8\times10^{-6}$ of $p$-value), are generally higher than those of local AGN. Using 929 X-ray selected AGN at $\langle z \rangle \sim 1.5$, \citet{Lusso12} showed that both types of AGN have higher Eddington ratio at higher redshift at any given $M_{\rm BH}$, which is in agreement with \citet{Netzer07} who used a sample of 9818 SDSS broad-line AGN at $z\leq0.75$. It should be noted that these literature adopted a different bolometric correction factors, spectral line fitting techniques estimating black hole mass (i.e., $L_{\rm Edd}$) from a different wavelength regime and redshift. Nevertheless, the observed trend of $\lambda_{\rm Edd}$ shown in Fig.~\ref{fig:histograms} is generally consistent with the literature.

\section{Discussion}
\label{sec:discussion}

In this section, we introduce caveat that should be taking into account when interpreting results, such as inconclusive correlation in the linear regression and selection bias. Then, we discuss role of mass-metallicity relation in the current sample. We also present positive correlations between \NV\ involved emission-line ratios and $\lambda_{\rm Edd}$ stating interpretations that help to understand observations. Lastly, we state lack of metallicity evolution of AGN with redshift.

The current samples used in this investigation are not enough to define the relation between $\lambda_{\rm Edd}$ and \NIIHa\ at high-redshift. As it is described in the Section~\ref{sec:ratio}, our sample of 53 AGN is neither complete nor fully representative of the high-redshift AGN. Nevertheless, we provide an observational hint that AGN beyond local Universe up to $z\sim1.7$ follow the same trend in $\lambda_{\rm Edd}$ and \NIIHa\ as low redshift AGN.

\subsection{Selection bias}
\begin{figure*}
\centering
	\includegraphics[width=0.60\textwidth]{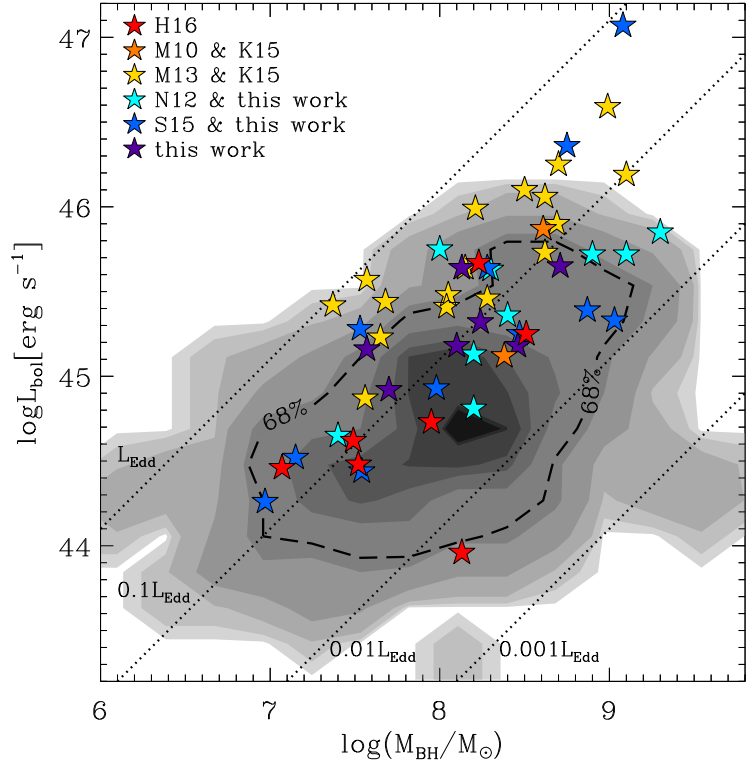}
    \caption{Bolometric luminosity vs. black hole mass for local ($\langle z \rangle \sim 0.036$, achromatic filled contours) hard X-ray selected AGN from \citet{Oh17} and high-z X-ray selected broad-line AGN ($\langle z \rangle \sim 1.327$, color-filled stars). The dotted lines represent the loci of Eddington ratios.}
    \label{fig:Lbol_Mbh}
\end{figure*}

\begin{figure*}
	\includegraphics[width=\linewidth]{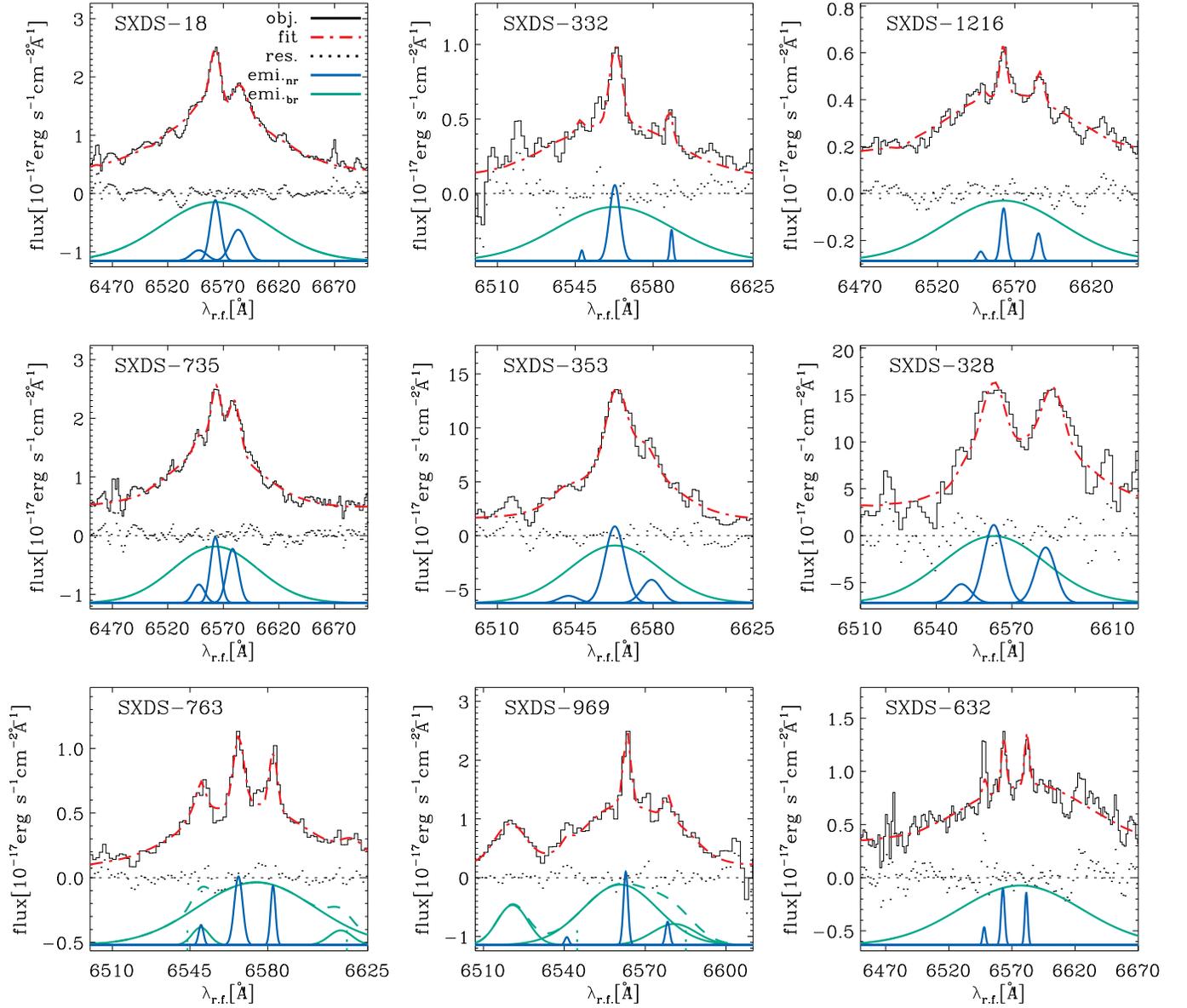}
    \caption{spectral line fit for 9 sources from \citet{Nobuta12}. Black lines present the observed spectra. Blue and green Gaussians indicate narrow and broad emission-line components, respectively. Red dash-dot line is the combined fit. Residuals are shown with black dots. 
    }
    \label{fig:fit_N12O18}
\end{figure*}

\begin{figure*}
	\includegraphics[width=\linewidth]{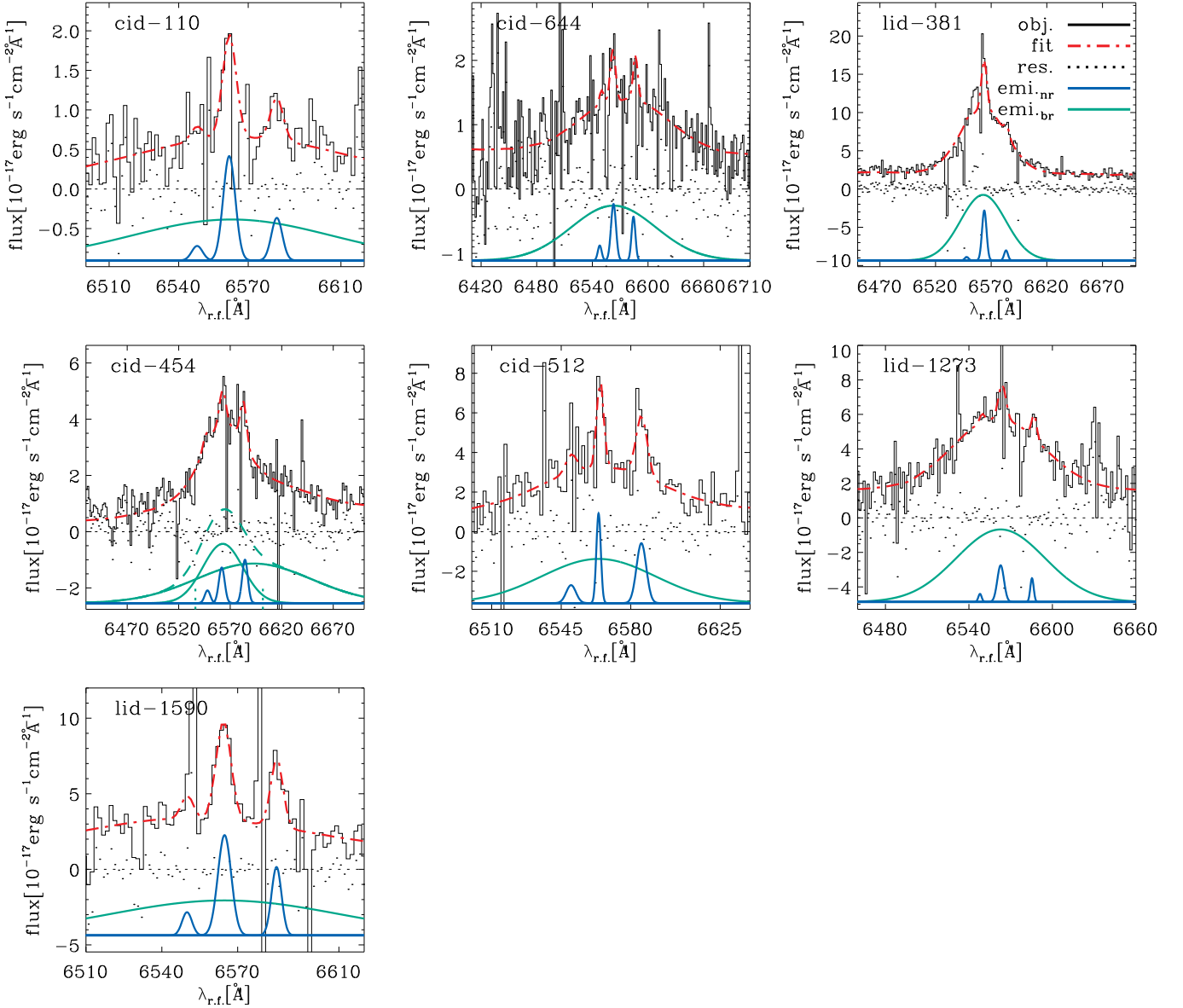}
    \caption{Spectral line fit for 7 sources from Suh et al. (\textit{in prep.}). The format is the same as that of Fig.~\ref{fig:fit_N12O18}. 
    }
    \label{fig:fit_S18O18}
\end{figure*}

\begin{figure*}
	\includegraphics[width=\linewidth]{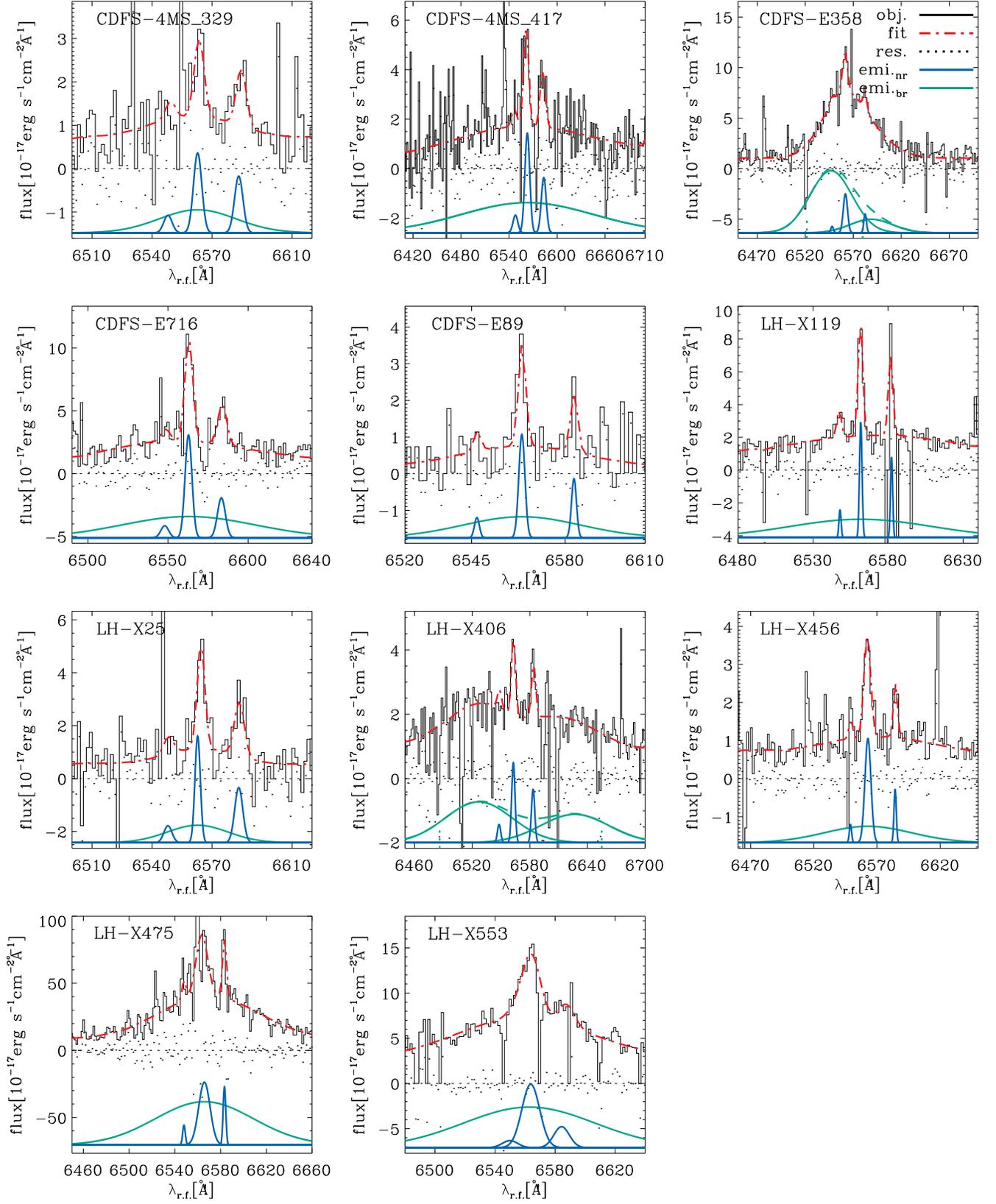}
    \caption{Spectral line fit for 11 sources from \citet{Suh15}. The format is the same as that of Fig.~\ref{fig:fit_N12O18}. 
    }
    \label{fig:fit_S15O18}
\end{figure*}

A possible selection bias could be present in our sample. The observability of the broad \Ha\ emission-line and narrow lines (\NII\ and \Ha) and following spectral line fitting are highly dependent on the line strengths and noise level of the spectra (i.e., S/N). Observational limitations caused by sensitivity of the instruments in such a high-redshift sample could lead us only to detect sources with prominent narrow emission-lines compared to the broad lines. Similarly, this is also true for $L_{\rm bol}$ that estimated from the X-ray surveys, which is not free from the survey flux limit. It should be noted that the primary flux limit is based on the ability to measure narrow emission-lines and  lower Eddington ratio sources need much more sensitive observations in the NIR. Fig.~\ref{fig:histograms} and Fig.~\ref{fig:Lbol_Mbh} present distribution of bolometric luminosity and Eddington ratio for local and high-z AGN. It is clearly seen that the high-z AGN presented in this work appear to be clumping in high $L_{\rm bol}$ and $\lambda_{\rm Edd}$.  

The distribution of the \NIIHa\ ratio reported in this paper and that of \citet{Oh17} is similar to that of the 12 \micron\ local AGN sample \citep{Malkan17}. The 12 \micron\ sample is representative of the local AGN population as it covers nearly six orders of magnitude in luminosity and because it is relatively unaffected by obscuration. \citet{Malkan17} investigated the optical and ultraviolet spectroscopic properties of 81 broad-line AGN and 104 narrow-line AGN and found the log(\NIIHa) ratio distribution has a median value of $-0.14$ compared to $-0.19$ (this work) and $-0.15$ \citep{Oh17}. The \NIIHa\ distribution of the IRAS 12 \micron\ AGN shows a somewhat broader distribution, compared to the X-ray selected local AGN and those in the present work, with a tail to low values of \NIIHa. The similar distributions of the \NIIHa\ ratio in X-ray and IR selected AGN imply that $\lambda_{\rm Edd}$ - \NIIHa\ relationship is probably not only confined to X-ray selected sample.

\subsection{Physical interpretations}
In \citet{Oh17} we discussed various complications such as the uncertainty in bolometric correction and AGN variability as a source of inherent scatter. We also discussed effect of mass-metallicity ($M_{*}-Z$) relation, X-ray heating, radiative-driven wind as possible mechanisms explaining the relationship between AGN Eddington ratio ($\lambda_{\rm Edd}$) and \NIIHa\ ratio (see \citealt{Oh17} for discussions and references therein). 

\citet{Larson74} predicted that metal abundances in galaxies depend on their stellar mass, as small galaxies experience large amount of gas loss via galactic winds. This picture was later observationally supported by \citet{Garnett02} and \citet{Tremonti04}. Narrow emission-line ratios in AGN have been reported to have a correlation with stellar mass as in star-forming galaxies \citep{Groves06, Stern13}. The same trend is also known for broad-line AGN \citep{Hamann99}. A positive relation between emission-line ratios and $M_{\rm BH}$ has been reported \citep{Shemmer02, Warner04, Nagao06a, Nagao06b, Matsuoka09, Matsuoka11, Oh17}. As it has been reported by \citet{Oh17}, however, not all emission-line ratios show statistically significant dependences on $M_{\rm BH}$ and $\lambda_{\rm Edd}$. If emission-line ratios and $\lambda_{\rm Edd}$ are tightly related due to $M_{*}-Z$ relation, other metallicity-sensitive emission-line ratios also should have shown the similar dependency. This implies that there are other important factors on the relationship between the emission-line ratios and $\lambda_{\rm Edd}$.

It should be noted that \citet{Matsuoka11} reported positive correlations between \NV\ involved emission-line ratios and $\lambda_{\rm Edd}$ using composite spectra of the SDSS quasars at $2.3<z<3.0$. Considering the fact that \NV\ broad emission-line is originated from regions much closer to the supermassive black hole in a sub-parsec scale, it is natural to depict \NV\ and \NII\ behave independently. As \citet{Matsuoka11} pointed out, composite spectra of luminous SDSS quasars seem to be associated with a post-starburst phase near the circumnuclear region that enriching nitrogen abundance by AGB stars, which is a different physical mechanism than that of the AGN samples explored in this work. Furthermore, distributions of $L_{\rm bol}$, $M_{\rm BH}$, and $\lambda_{\rm Edd}$ for the 2678 SDSS quasars used in their work show that these quasars exhibit more than an order of magnitude larger $L_{\rm bol}$, $M_{\rm BH}$, and $\lambda_{\rm Edd}$ than those presented in this work. As their samples are observed in higher redshift with intermediate spectral resolution (R$\sim2000$) optical spectrograph, bright quasars are preferentially selected, which in turn showing remarkably luminous, massive and high Eddington ratio quasars. As a result, baseline of $\lambda_{\rm Edd}$ is partially restricted from $-1.1$ to $-0.1$. On the other hand, samples used in \citet{Oh17} and this work are X-ray selected AGN in a lower redshift ($ \langle z \rangle = 0.036$ and $1.327$, respectively) representing the majority of AGN populations with wide ranges of $L_{\rm bol}$ ($\sim3.5$ dex), $M_{\rm BH}$ ($\sim4$ dex), and $\lambda_{\rm Edd}$ ($\sim4$ dex). Therefore, we interpret the anti-correlated observational relation in \NIIHa\ vs. $\lambda_{\rm Edd}$ as a global feature appearing in narrow-line region of typical AGN at $z<1.7$.

In this work, we do not find evidence of metallicity evolution of AGN with the redshift. The \NIIHa\ ratio, which is often used as a metallicity indicator, and its distribution of local and high-z AGN shown in Fig.~\ref{fig:ratio} appear similar trend in a limited range of ${\rm log}\lambda_{\rm Edd}$ (at $>-1.81$) as mentioned in Section~\ref{sec:ratio}. This is consistent with previous studies, which presented that the gas-phase metallicity of AGN has been shown to be constant in super-solar across the majority of cosmic time unlike inactive galaxies \citep{Dietrich03, Nagao06b, Juarez09, Matsuoka09, DorsJr14}.

\section{Summary}
We have presented observational results on the relationship between AGN Eddington ratio ($\lambda_{\rm Edd}$) and narrow emission line ratio (\NIIHa) using 53 X-ray selected broad-line AGN at $0.6<z<1.7$ compiled from the literature and the spectroscopic observations. Our main findings are as follows. 

\begin{itemize}
	\item The distribution of \NIIHa\ for X-ray selected broad-line AGN at $0.6<z<1.7$ as a function of $\lambda_{\rm Edd}$ is similar with those of local AGN at $ \langle z \rangle = 0.036$, providing an observational clue that \NIIHa\ vs. $\lambda_{\rm Edd}$ relation still holds well up to the given redshift. However, it is difficult to draw a statistically definite conclusion on the linear regression analysis, due to paucity of low $\lambda_{\rm Edd}$ sources in high-z regime. 	
	\item The observed trend between $\lambda_{\rm Edd}$ and \NIIHa\ ratio could be explained by considering the mass-metallicity relation, X-ray heating processes, and prevalent radiatively driven outflows that depend on the $\lambda_{\rm Edd}$ state. 
	\item Metallicity evolution of AGN is not observed, which is consistent with the literatures.
\end{itemize}

\tabletypesize{\scriptsize}
\begin{deluxetable*}{lccccccc}
\tablecaption{Properties of High-redshift Broad-Line AGN\label{tab:data}}
\tablewidth{0pt}
\tablecolumns{8}
\tablehead{
\colhead{Source\tablenotemark{a}} &
\colhead{Field\tablenotemark{b}} &
\colhead{ID\tablenotemark{c}} &
\colhead{z} &
\colhead{log(\NIIHa)} & 
\colhead{log$L_{\rm bol}$\tablenotemark{d}} &
\colhead{log$(M_{\rm BH}/M_{\rm \odot})$} &
\colhead{log$\lambda_{\rm Edd}$} 
}
\startdata
  	   KM10&       CCLS&        543&      1.300&                       $-0.46$  &                       $45.12$  &                        $8.38$  &                      $-1.37$\\
       KM10&       CCLS&       1930&      1.565&                       $-0.08$  &                       $45.87$  &                        $8.61$  &                      $-0.86$\\
       KM13&       CCLS&        142&      0.699&                       $-0.06$  &                       $45.41$  &                        $8.04$  &                      $-0.75$\\
       KM13&       CCLS&        192&      1.222&                       $-0.18$  &                       $45.44$  &                        $7.68$  &                      $-0.35$\\
       KM13&       CCLS&        208&      1.242&                       $-0.11$  &                       $46.19$  &                        $9.10$  &                      $-1.02$\\
       KM13&       CCLS&        607&      1.297&                       $-0.22$  &                       $45.99$  &                        $8.21$  &                      $-0.33$\\
       KM13&       CCLS&       1194&      1.314&                       $-0.16$  &                       $44.87$  &                        $7.56$  &                      $-0.80$\\
       KM13&       CCLS&        549&      1.319&                       $-0.19$  &                       $46.59$  &                        $8.99$  &                      $-0.51$\\
       KM13&       CCLS&        112&      1.322&                       $-0.08$  &                       $46.06$  &                        $8.62$  &                      $-0.68$\\
       KM13&       CCLS&        463&      1.327&                        $0.08$  &                       $45.42$  &                        $7.37$  &                      $-0.07$\\
       KM13&       CCLS&        157&      1.332&                       $-0.10$  &                       $45.46$  &                        $8.28$  &                      $-0.94$\\
       KM13&       CCLS&        604&      1.345&                       $-0.41$  &                       $45.57$  &                        $7.57$  &                      $-0.12$\\
       KM13&       CCLS&        546&      1.407&                       $-0.00$  &                       $45.90$  &                        $8.69$  &                      $-0.90$\\
       KM13&       CCLS&        499&      1.454&                       $-0.11$  &                       $46.25$  &                        $8.70$  &                      $-0.57$\\
       KM13&       CCLS&        119&      1.505&                       $-0.17$  &                       $45.48$  &                        $8.05$  &                      $-0.69$\\
       KM13&       CCLS&       1044&      1.561&                        $0.19$  &                       $46.10$  &                        $8.50$  &                      $-0.52$\\
       KM13&       CCLS&        216&      1.567&                       $-0.09$  &                       $45.23$  &                        $7.65$  &                      $-0.54$\\
       KM13&       CCLS&        305&      1.572&                       $-0.32$  &                       $45.73$  &                        $8.62$  &                      $-1.00$\\
       KM13&       CCLS&        255&      1.667&                       $-0.32$  &                       $45.64$  &                        $8.15$  &                      $-0.62$\\
        N12&       SXDS&        328&      0.809&                       $-0.15$  &                       $44.81$  &                        $8.20$  &                      $-1.50$\\
        N12&       SXDS&        353&      0.989&                       $-0.52$  &                       $44.65$  &                        $7.40$  &                      $-0.86$\\
        N12&       SXDS&        332&      1.385&                       $-0.64$  &                       $45.63$  &                        $8.30$  &                      $-0.78$\\
        N12&       SXDS&        763&      1.413&                       $-0.19$  &                       $45.75$  &                        $8.00$  &                      $-0.36$\\
        N12&       SXDS&        735&      1.447&                       $-0.01$  &                       $45.36$  &                        $8.40$  &                      $-1.15$\\
        N12&       SXDS&         18&      1.452&                       $-0.12$  &                       $45.72$  &                        $8.90$  &                      $-1.29$\\
        N12&       SXDS&       1216&      1.472&                       $-0.28$  &                       $45.85$  &                        $9.30$  &                      $-1.56$\\
        N12&       SXDS&        969&      1.585&                       $-0.32$  &                       $45.13$  &                        $8.20$  &                      $-1.18$\\
        N12&       SXDS&        632&      1.593&                       $-0.06$  &                       $45.72$  &                        $9.10$  &                      $-1.49$\\
        S15&       CDFS&        716&      0.763&                       $-0.37$  &                       $44.93$  &                        $7.98$  &                      $-1.16$\\
        S15&         LH&        456&      0.877&                       $-0.52$  &                       $45.28$  &                        $7.53$  &                      $-0.36$\\
        S15&       CDFS&        329&      0.954&                       $-0.15$  &                       $44.26$  &                        $6.97$  &                      $-0.82$\\
        S15&         LH&        475&      1.205&                       $-0.49$  &                       $47.07$  &                        $9.08$  &                      $-0.12$\\
        S15&       CDFS&        417&      1.222&                       $-0.26$  &                       $45.33$  &                        $9.03$  &                      $-1.81$\\
        S15&         LH&        406&      1.283&                       $-0.17$  &                       $45.39$  &                        $8.87$  &                      $-1.59$\\
        S15&         LH&        119&      1.406&                       $-0.17$  &                       $45.25$  &                        $8.47$  &                      $-1.33$\\
        S15&         LH&        553&      1.440&                       $-0.54$  &                       $46.36$  &                        $8.75$  &                      $-0.50$\\
        S15&         LH&         25&      1.599&                       $-0.18$  &                       $44.52$  &                        $7.15$  &                      $-0.74$\\
        S15&       CDFS&         89&      1.613&                       $-0.24$  &                       $44.44$  &                        $7.54$  &                      $-1.21$\\
        S15&       CDFS&        358&      1.626&                       $-0.44$  &                       $45.64$  &                        $8.28$  &                      $-0.75$\\
        H16&       SXDS&        194&      0.627&                       $-0.94$  &                       $44.48$  &                        $7.52$  &                      $-1.15$\\
        H16&       SXDS&        827&      0.658&                       $-0.30$  &                       $44.62$  &                        $7.49$  &                      $-0.99$\\
        H16&       CDFS&        629&      0.667&                        $0.08$  &                       $43.96$  &                        $8.13$  &                      $-2.29$\\
        H16&       SXDS&        393&      0.822&                       $-0.28$  &                       $44.73$  &                        $7.95$  &                      $-1.34$\\
        H16&       SXDS&        600&      0.873&                       $-0.26$  &                       $44.46$  &                        $7.07$  &                      $-0.73$\\
        H16&       SXDS&        883&      0.961&                        $0.26$  &                       $45.67$  &                        $8.23$  &                      $-0.67$\\
        H16&       CDFS&        101&      0.977&                       $-0.02$  &                       $45.25$  &                        $8.51$  &                      $-1.37$\\
        S19&       CCLS&        110&      0.729&                       $-0.42$  &                       $44.92$  &                        $7.70$  &                      $-0.89$\\
        S19&       CCLS&        381&      0.767&                       $-0.74$  &                       $45.16$  &                        $7.57$  &                      $-0.52$\\
        S19&       CCLS&        644&      0.986&                       $-0.19$  &                       $45.18$  &                        $8.10$  &                      $-1.03$\\
        S19&       CCLS&        454&      1.485&                        $0.07$  &                       $45.32$  &                        $8.24$  &                      $-1.04$\\
        S19&       CCLS&        512&      1.516&                        $0.03$  &                       $45.64$  &                        $8.13$  &                      $-0.60$\\
        S19&       CCLS&       1590&      1.596&                       $-0.26$  &                       $45.65$  &                        $8.71$  &                      $-1.17$\\
        S19&       CCLS&       1273&      1.622&                       $-0.37$  &                       $45.18$  &                        $8.46$  &                      $-1.39$\\             
\enddata
\tablenotetext{a}{Source of literature: KM10 \citep{Kartaltepe15, Merloni10}; KM13 \citep{Kartaltepe15, Matsuoka13}; N12 \citep{Nobuta12}; S15 \citep{Suh15}; H16 \citep{Harrison16}; S19 (Suh et al. in prep.).}
\tablenotetext{b}{Field name that used to measure $2-10$ keV luminosity: CCLS \citep{Civano16}; SXDS \citep{Ueda08}; CDFS \citep{Xue11}; LH \citep{Brunner08}.}
\tablenotetext{c}{X-ray source ID from the literature.}
\tablenotetext{d}{In units of \ergs.}
\end{deluxetable*}

\acknowledgments
We thank Richard Mushotzky for taking the time to give valuable advice. 
K.O. acknowledges support from the Japan Society for the Promotion of Science (JSPS, ID: 17321). Part of this work was financially supported by the Grant-in-Aid for Scientific Research 17K05384 (Y.U.).
M.K. acknowledges support from NASA through ADAP award NNH16CT03C.  

This research has made use of NASA's ADS Service. 
This research has made use of the NASA/ IPAC Infrared Science Archive, which is operated by the Jet Propulsion Laboratory, California Institute of Technology, under contract with the National Aeronautics and Space Administration.

%

\vspace{5mm}
\facilities{Subaru(FMOS), IRSA}


\software{gandalf \citep{Sarzi06}}

\bibliography{ms}



\end{document}